\documentclass[aps,prc,reprint,groupedaddress]{revtex4-1}
\usepackage{graphicx}

\begin{document}


\title{Effects of consistent and inconsistent isobar coupling in the nuclear medium}

\author{C. L. Korpa}
\affiliation{Department of Theoretical physics, University of P\'ecs, Ifj\'us\'ag \'utja 6, 7624 P\'ecs, Hungary}

\date{\today}

\begin{abstract}
We investigate effects of consistent (conserving the number of degrees of freedom)
and inconsistent pion-nucleon-isobar couplings on the isobar propagator in vacuum and
in nuclear medium. Using the consistent coupling in conjunction with a convenient basis 
leads to significant simplification of the isobar vacuum and in-medium self energy and 
dressed propagator compared to the case of inconsistent interaction. The higher-derivative 
nature of the consistent interaction requires a suitable compensation by an additional
form-factor term or adjustment in the used cut-off values. This modification is 
straightforward to perform and assures that the 
physical observables connected to the spin-3/2 sector in vacuum and nuclear medium acquire 
values indistinguishable from the ones obtained by using an inconsistent coupling.
\end{abstract}

\pacs{24.10.Jv,25.20.Dc,21.65.-f,13.75.Gx}

\maketitle

\section{Introduction \label{Intro}}
Relativistically covariant description of particles with spin one or higher 
faces the problem of too many degrees of freedom present in the covariant
field variables. The superfluous degrees of freedom can be eliminated
by imposing auxiliary conditions as, for example, in using the Rarita-Schwinger 
description \cite{RS-1941} of the spin-3/2 isospin-3/2 $\Delta(1232)$ baryon
(in the following referred to as the isobar). However, after introducing 
interactions of the high-spin fields in general they acquire unphysical 
degrees of freedom, for example in the isobar propagator spin-1/2 components
appear. A way out of this problem has been shown in Ref.\ \cite{Pascalutsa-1998}, 
where couplings consistent with the gauge invariance of the Rarita-Schwinger field 
were constructed, which preserved the correct number of degrees of freedom.
Further investigations have shone more light on the fundamental and phenomenological
relations between the widely used inconsistent (in the sense of not conserving the
number of degrees of freedom) and gauge invariant (consistent) interactions
\cite{PT-1999, Pascalutsa-2001, Pascalutsa-2008}.

In general the consistent coupling of a higher-spin field is of higher order
in derivatives than the inconsistent interaction with the lowest number of 
derivatives. Recently it has been pointed out \cite{Vrancx-2011a,Vrancx-2011b} 
that this introduces additional incoming-energy powers in the cross sections 
which could lead to incorrect determination of the mass of high-spin intermediate
particle because of the shifted energy peak. In order to avoid this problem a 
suitable modification of an applied form-factor was suggested \cite{Vrancx-2011a}.

\section{Isobar propagator in vacuum and pion-nucleon scattering \label{s1}}
We take the free isobar propagator in the standard form
\begin{eqnarray}
G_0^{\mu\nu}(p)&=&\frac{p\hspace{-1.6mm}/ +M_\Delta}{p^2-M^2_\Delta+
i\varepsilon}
\left[ g^{\mu\nu}-\frac{\gamma^\mu\gamma^\nu}{3}
-\frac{2p^\mu p^\nu}{3M^2_\Delta} \right. \nonumber \\
&+& \left. \frac{p^\mu\gamma^\nu-p^\nu\gamma^\mu}{3M_\Delta}
\right].\label{freeprop}
\end{eqnarray}
and start by computing the full relativistically covariant structure of the
isobar propagator in vacuum using inconsistent as well as consistent pion-nucleon-isobar
couplings. For the former we take the widely used form \cite{NEK-1971}:
\begin{equation}
{\mathcal{L}}_{\pi\text{N}\Delta}=
g_{\pi\text{N}\Delta} \partial_\alpha \pi \bar{\Delta}_\beta
(g^{\alpha\beta}+a \gamma^\beta \gamma^\alpha)N + \text{h.c.}
\label{pindinc}
\end{equation}
with the off-shell parameter $a$. For the consistent coupling we take the nontrivial 
lowest order in derivatives expression \cite{Pascalutsa-1998}:
\begin{equation}
{\mathcal{L}}^{\text{(con)}}_{\pi\text{N}\Delta}=
\frac{g^{\text{(con)}}_{\pi\text{N}\Delta}}{M_\Delta}\varepsilon^{\mu\nu\alpha\beta} \partial_\mu \bar{\Delta}_\nu
\gamma_5 \gamma_\alpha N\,\partial_\beta \pi + \text{h.c.},
\label{pindcon}
\end{equation}
with isospin factors not shown in both cases.

To represent the isobar self-energy and propagator we use a convenient orthogonal basis introduced in Ref.\ 
\cite{LK-2002} and used for describing the isobar in the nuclear medium in Ref.\ \cite{KD-2004}. 
The basis consists of 40 terms made up by $Q^{\mu\nu}_{[ij]}$ with $i,j=1,2$ and $P^{\mu\nu}_{[ij]}$
with $i,j=1,\dots ,6$ which depend on the isobar four-momentum and the four-velocity of the nuclear medium and 
satisfy the orthogonality relations (for defintion and details see Ref.~\cite{KD-2004}): 
\begin{eqnarray}
&& Q_{[ik]}^{\mu \alpha }\,g_{\alpha \beta}\,P_{[lj]}^{\beta \nu }
= 0 = P_{[ik]}^{\mu \alpha }\,g_{\alpha \beta}\,Q_{[lj]}^{\beta \nu }
\;, \nonumber\\
&& Q_{[ik]}^{\mu \alpha }\,g_{\alpha \beta}\,Q_{[lj]}^{\beta \nu }
= \delta_{kl}\,Q_{[ij]}^{\mu \nu} \;,\quad 
P_{[ik]}^{\mu \alpha }\,g_{\alpha \beta}\,P_{[lj]}^{\beta \nu }
= \delta_{kl}\,P_{[ij]}^{\mu \nu}\;.
\label{orthog}
\end{eqnarray}
In vacuum only those elements are relevant which do not contain the four-velocity of the 
medium. These elements can be obtained by defining
\begin{eqnarray}
Q'^{\mu\nu}_{[11]}&=& Q^{\mu\nu}_{[11]}+P^{\mu\nu}_{[55]}\nonumber \\ 
Q'^{\mu\nu}_{[22]}&=& Q^{\mu\nu}_{[22]}+P^{\mu\nu}_{[66]}, \label{qprime}
\end{eqnarray}
and taking the diagonal terms $P^{\mu\nu}_{[11]},P^{\mu\nu}_{[22]},P^{\mu\nu}_{[33]},P^{\mu\nu}_{[44]}$
as well as the non-diagonal ones: $P^{\mu\nu}_{[13]},P^{\mu\nu}_{[31]},P^{\mu\nu}_{[24]},P^{\mu\nu}_{[42]}$
providing the required 10 terms already used, in different form, in Ref.~\cite{Korpa-1997fk}. 
This is in accordance with the presence, apart from the spin-3/2 part, of two spin-1/2 sectors of opposite
parity and their mixing in the vacuum Rarita-Schwinger propagator \cite{KL-2006}.
The spin-3/2 and spin-1/2 projection operators \cite{BDM-1989}  can be expressed as:
\begin{eqnarray}
(P^{3/2})^{\mu\nu}&=& Q'^{\mu\nu}_{[11]}+Q'^{\mu\nu}_{[22]}, \nonumber \\ 
(P^{1/2}_{11})^{\mu\nu}&=& P^{\mu\nu}_{[11]}+P^{\mu\nu}_{[22]},\nonumber \\ 
(P^{1/2}_{22})^{\mu\nu}&=& P^{\mu\nu}_{[33]}+P^{\mu\nu}_{[44]},\nonumber \\
(P^{1/2}_{12})^{\mu\nu}&=& P^{\mu\nu}_{[13]}+P^{\mu\nu}_{[31]}\nonumber \\ 
(P^{1/2}_{21})^{\mu\nu}&=& P^{\mu\nu}_{[24]}+P^{\mu\nu}_{[42]}. \label{pspacevac}
\end{eqnarray}
The Lorentz structure of the inconsistent coupling (\ref{pindinc}) can be similarly
decomposed (in momentum space):
\begin{eqnarray}
(g^{\mu\nu}+a\gamma^\mu\gamma^\nu)&q_\nu &= \left( 
Q'^{\mu\nu}_{[11]}+Q'^{\mu\nu}_{[22]}+(1+3a)(P^{\mu\nu}_{[11]}+P^{\mu\nu}_{[22]})\right. \nonumber\\ 
&+& (1+a)(P^{\mu\nu}_{[33]}+P^{\mu\nu}_{[44]})\nonumber \\
&+&\sqrt{3}a(P^{\mu\nu}_{[13]}+P^{\mu\nu}_{[31]}-P^{\mu\nu}_{[24]}
- \left. P^{\mu\nu}_{[42]})\right) q_\nu,
\end{eqnarray}
with $q$ being the pion four-momentum. The expression for the consistent coupling (\ref{pindcon})
turns out much simpler in this basis:
\begin{equation}
\varepsilon^{\mu\alpha\beta\nu} \gamma_5 \gamma_\alpha p_\beta q_\nu=
\sqrt{p^2}\left( Q'^{\mu\nu}_{[11]}-Q'^{\mu\nu}_{[22]}+2(P^{\mu\nu}_{[11]}-P^{\mu\nu}_{[22]})
\right) q_\nu,
\end{equation}
where $p$ is the isobar four-momentum.

\begin{figure}
\hspace{-3mm}
\includegraphics[width=9.5cm]{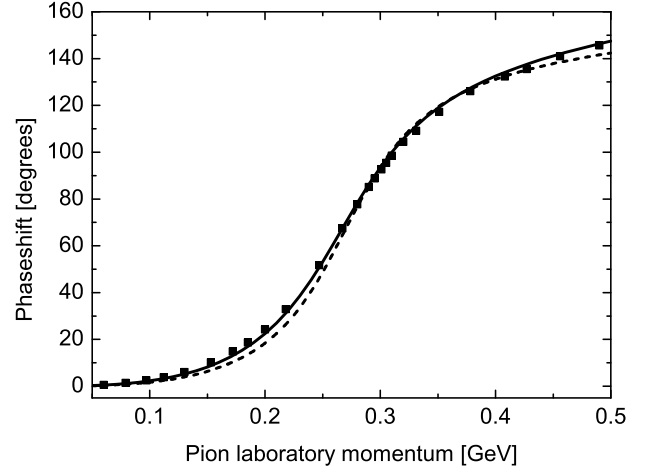}
\caption{Pion-nucleon phaseshift in the spin-3/2 isospin-3/2 channel calculated with the consistent
coupling (\ref{pindcon}). The points represent 
measurements from Ref.\ \cite{KP-1980}, the dash line is the calculated result with 
coupling and form-factor from the fit using the inconsistent coupling (\ref{pindinc}), 
while the solid line is obtained with insertion of additional form-factor (\ref{ffadd}).
\label{phasesh}}
\end{figure}

Expanding the free isobar propagator (and its inverse) in terms of $Q'^{\mu\nu}_{[ij]}$ and
$P^{\mu\nu}_{[ij]}$ and doing the same with the vacuum self energy:
\begin{equation}
\Sigma_V^{\mu\nu}(p)=\sum_{i,j=1}^2 Q'^{\mu\nu}_{[ij]}\, \sigma^{(Q')}_{V[ij]}(p)+
\sum_{i,j=1}^4 P^{\mu\nu}_{[ij]} \,\sigma^{(P)}_{V[ij]}(p),\label{vacself}
\end{equation}
the Schwinger-Dyson equation for the vacuum propagator
\begin{equation}
G_V^{\mu\nu}=G_0^{\mu\nu}+G_0^{\mu\alpha}\,g_{\alpha\beta}\,\Sigma_V^{\beta\gamma}\,
g_{\gamma\delta}\,G_V^{\delta\nu}
\end{equation}
becomes a matrix equation for the coefficients appearing in the expansion of the dressed vacuum propagator:
\begin{equation}
G_V^{\mu\nu}(p)=\sum_{i,j=1}^2 Q'^{\mu\nu}_{[ij]}\, G^{(Q')}_{V[ij]}(p)+
\sum_{i,j=1}^4 P^{\mu\nu}_{[ij]} \,G^{(P)}_{V[ij]}(p)\label{vacprop}.
\end{equation}
The obtained propagator
we use to repeat the calculation of Ref.~\cite{Korpa-1997fk} for the pion-nucleon
phaseshift in the spin-3/2 isospin-3/2 channel, but using the consistent coupling (\ref{pindcon}). The 
isobar self energy for this case is given in Appendix by expressions (\ref{convacself}).
In the adopted basis the computation is especially simple, since the phaseshift $\delta$ is given through the ratio of 
the imaginary and real part of
the coefficient of $Q'^{\mu\nu}_{[11]}$ since this projector corresponds to the positive-energy spin-3/2 term:
\begin{equation}
\delta=\arctan\,\frac{\text{Im}\,G^{(Q')}_{V[11]}(p)}{\text{Re}\,G^{(Q')}_{V[11]}(p)}.
\end{equation} 
In Fig.~\ref{phasesh} the obtained phaseshift is shown by dash line using parameter values form 
Ref.~\cite{Korpa-1997fk}, i.e.\ $g^{\text{(con)}}_{\pi\text{N}\Delta}/M_\Delta=19\,\text{GeV}^{-1}$
and $\Lambda=0.97\,\text{GeV}$ in the form-factor
\begin{equation}
F(p^2)=\exp\left[-\frac{p^2-(M_N+m_\pi)^2}{\Lambda^2}\right].\label{vacuumff}
\end{equation}
The higher derivative order of the consistent coupling introduces a slight departure from the
result with the inconsistent coupling seen as a small mismatch of the dash line and measurement points, 
but this can be easily corrected by an additional term 
in the form-factor, as advocated in Refs.~\cite{Vrancx-2011a,Vrancx-2011b}:
\begin{equation}
F_c(p^2)=\left( \frac{\Lambda_c^2}{p^2-M_\Delta^2+\Lambda_c^2}\right)^{1/2}.
\label{ffadd}
\end{equation}
Using $\Lambda_c=1.2\,\text{GeV}$ the solid-line result in Fig.~\ref{phasesh} is obtained. The 
non-analytic form of the expression (\ref{ffadd}) is not expected to present any problems in physical 
applications where always the square of the form-factor appears. Also, the pole in expression (\ref{ffadd})
should be kept far from the
physical region, i.e.\ one should use $\Lambda_c\approx M_\Delta$.

\begin{figure}
\includegraphics[width=9.5cm]{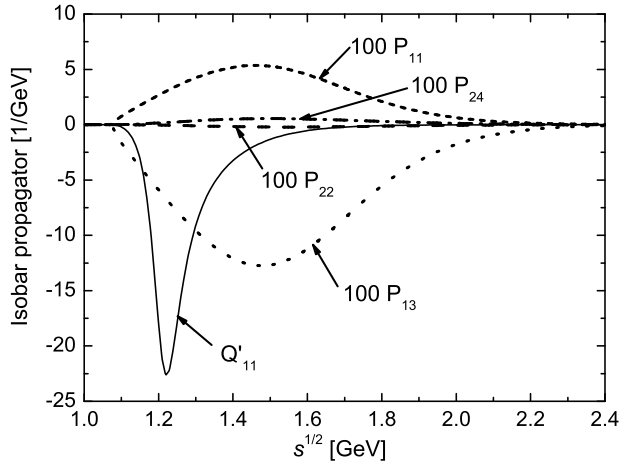}
\caption{Imaginary parts of the isobar vacuum propagator components for inconsistent coupling. The 
components of the spin-1/2 sector have been multiplied by 100 in order to be visible.
\label{vacpropfig1}}
\end{figure}

\begin{figure}
\includegraphics[width=9.5cm]{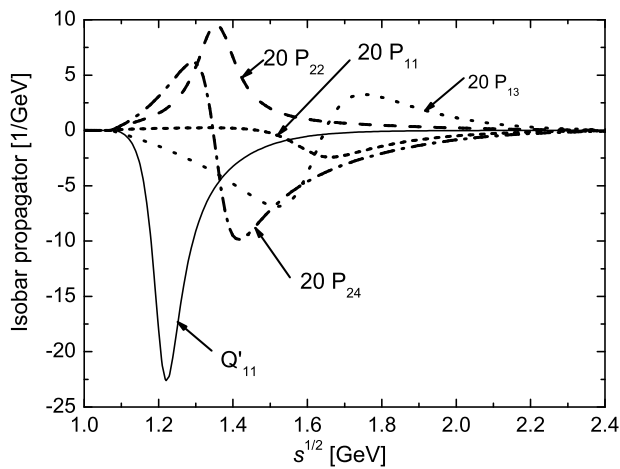}
\caption{The same as Fig.~\ref{vacpropfig1} but when the off-shell parameter $a=-1$, with
multiplication factor 20 for the spin-1/2 terms.
\label{vacpropfig2}}
\end{figure}

The vacuum self-energy (\ref{incvacself}) for inconsistent coupling has many more nonzero
components than the one for the consistent coupling (\ref{convacself}) and they induce  
nonzero vacuum propagator components which are zero for consistent coupling. However, they are
in general much smaller than the dominant spin-3/2 component, as can be seen in Fig.~\ref{vacpropfig1}
which shows the imaginary parts of the vacuum propagator components (coefficients of projectors
$Q'^{\mu\nu}_{[11]}$ and $P^{\mu\nu}_{[ij]}$) when the off-shell parameter $a=0$. 
The terms corresponding to the spin-1/2 sector have
been multiplied by 100.
Choosing $a=-1$ produces significantly different spin-1/2 sector as shown in Fig.~\ref{vacpropfig2},
but these terms are still much smaller than the spin-3/2 sector. In order to be able to see them 
they have been multiplied by a factor of 20 in Fig.~\ref{vacpropfig2}.
One can conclude that the terms induced in the isobar propagator by the inconsistent coupling (\ref{pindinc}) in the 
spin-1/2 sector are much smaller (by about two oreders of magnitude) from the spin-3/2 contribution if the
off-shell parameter is of order unity. While this effect gives negligible contributions to the intrinsically
spin-3/2 related observables it can have non-negligible effects when fitting observables related to the
spin-1/2 background thus affecting the obtained parameter set characterizing the interactions. This is not
surprising since it has been shown \cite{Pascalutsa-2001} that by field  redefintion an inconsistent isobar
coupling can be transformed into a consistent one plus additional interactions involving the spin-1/2 degrees 
of freedom. With the latter approach a clear separation of different contributions to observables concerning their origin 
is achieved.
\section{Isobar propagator in the nuclear medium and nuclear photoabsorption \label{s2}}
We turn to the investigation of the in-medium isobar propagator. In this case, even in 
rotatioanlly symmetric nuclear medium, the rotational symmetry for a moving isobar is
broken, since the direction of the velocity with respect to the medium introduces a 
preferred direction and only rotations around that axis are symmetry operations
\cite{KL-2003}. This means that only the helicity is a good quantum number and the spin-1/2
sector gets legitimate contributions even for the case of consistent isobar coupling.
The projectors $Q^{\mu\nu}_{[ij]}$ with $i,j=1,2$ correspond to the helicity-3/2 space,
while the $P^{\mu\nu}_{[ij]}$ with $i,j=1,\dots,6$ span the helicity-1/2 space.
In this way the helicity degeneracy in vacuum is lifted and the propagator coefficients
of $Q^{\mu\nu}_{[ii]}$ and $P^{\mu\nu}_{[4+i,4+i]}$ ($i=1,2$) are not identical if the isobar
is moving in the medium. 

\begin{figure}
\includegraphics[width=9.5cm]{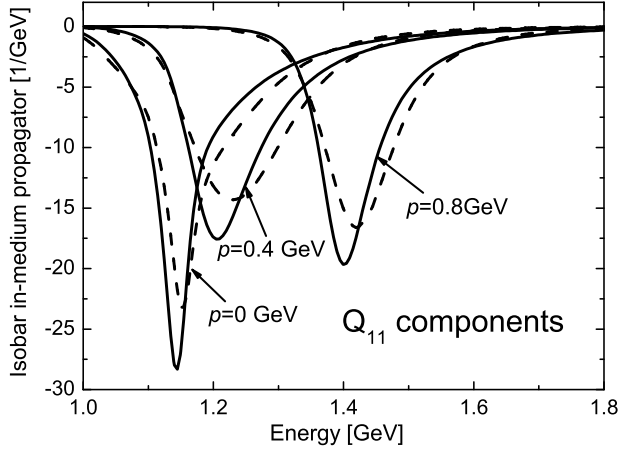}
\caption{Imaginary part of the $G^{(Q)}_{[11]}(p_0,|\vec p|)$ coefficient in the isobar propagator expansion 
(\ref{medprop}). Solid line is for the consistent pion-nucleon-isobar coupling and dash line for the 
inconsistent coupling. The two sets of curves coincide if the effect of the additional form-factor 
(\ref{ffadd}) is included for the consistent coupling with $\Lambda_c=1.2$GeV.
\label{medprop-1}}
\end{figure}

\begin{figure}
\includegraphics[width=9.5cm]{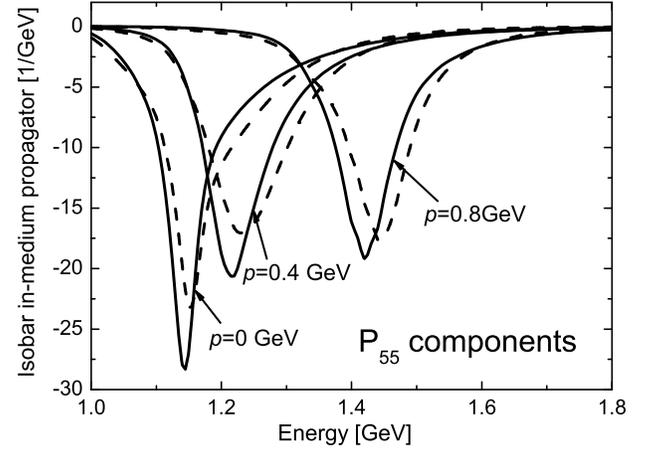}
\caption{The same as Fig.~\ref{medprop-1} but for the imaginary part of $G^{(P)}_{[55]}(p_0,|\vec p|)$. 
\label{medprop-2}}
\end{figure}

The isobar in-medium self-energy decomposition now becomes:
\begin{equation}
\Sigma^{\mu\nu}(p,u)=\sum_{i,j=1}^2 Q^{\mu\nu}_{[ij]}\, \sigma^{(Q)}_{[ij]}(p,u)+
\sum_{i,j=1}^6 P^{\mu\nu}_{[ij]} \,\sigma^{(P)}_{[ij]}(p,u),\label{medself}
\end{equation}
with expansion coefficients $\sigma^{(Q)}_{[ij]}$ and $\sigma^{(P)}_{[ij]}$ given
in Appendix based on the pion-nucleon loop. The isobar propagator is expanded in an
analogous way:
\begin{equation}
G^{\mu\nu}(p,u)=\sum_{i,j=1}^2 Q^{\mu\nu}_{[ij]}\, G^{(Q)}_{[ij]}(p,u)+
\sum_{i,j=1}^6 P^{\mu\nu}_{[ij]} \,G^{(P)}_{[ij]}(p,u).\label{medprop}
\end{equation}
We present results for both consistent 
and inconsistent coupling and including mean-field shifts of the nucleon mass and energy
in the medium. In order to suppress the pion off-shell contribution we use an off-shell form
factor of the dipole form:
\begin{equation}
F_\pi(q_\pi)=\left( \frac{ \Lambda^2_\pi-m^2_\pi}{ \Lambda^2_\pi-q^2_\pi} \right)^2,
\label{medff}
\end{equation}
with $\Lambda_\pi=1.2$GeV. The in-medium pion propagator we take from Ref.~\cite{KLR-2009}
and use the same values for the nucleon mean-field self energy as in Ref.~\cite{RLK-2009}. We concentrate on comparing
results for the nuclear photoabsorption in the isobar region and thus perform calculations at
0.8 times saturation density by taking the Fermi momentum $k_F=250$MeV. The used nucleon mean-field
shifts are obtained by rescaling the saturation-density values \cite{RLK-2009} giving 
$\Sigma_N^s=-0.28$GeV and $\Sigma_N^v=0.232$GeV for the scalar and vector contributions. 
We first examine the dominant contributions to the isobar
propagator by comparing the imaginary parts of the $Q_{[11]}^{\mu\nu}$ and $P_{[55]}^{\mu\nu}$ projectors. They
correspond to (on-shell) positive energy helicity-3/2 and helicity-1/2 terms originating from the vacuum spin-3/2
sector. We work in the rest frame of the medium and use the isobar energy $p_0$ and momentum $|\vec p|$ as
variables. In Fig.~\ref{medprop-1} we show the helicity-3/2 components ($Q^{\mu\nu}_{[11]}$ coefficients)
for diferent isobar momenta, with solid line corresponding to consistent coupling and dash line to inconsistent
one using identical form-factors. Fig.~\ref{medprop-2} shows analogous helicity-1/2 results, i.e.\ coefficients 
of the $P_{[55]}^{\mu\nu}$ projector.
Including the effect of the additional form-factor (\ref{ffadd}) with $\Lambda_c=1.2$GeV for the
consistent coupling gives curves indistinguishable from the ones of the inconsistent coupling. 
 
\begin{figure}
\includegraphics[width=9.5cm]{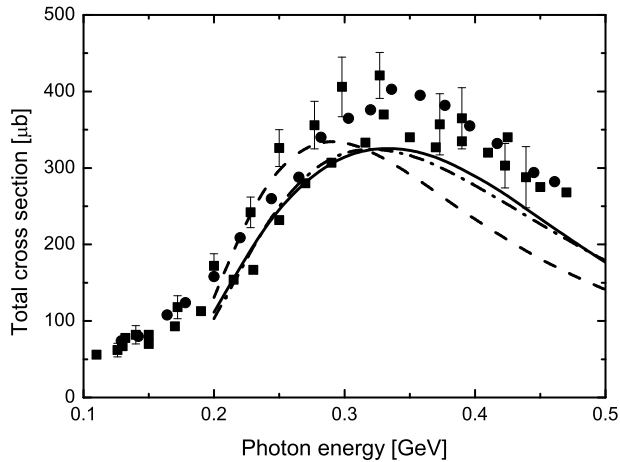}
\caption{Calculated nuclear photoabsorption cross section (lines) compared to measurements on heavy
nuclei (points with and without error bars) from Ref.~\cite{Ahrens-1984}. Solid line is for the 
inconsistent and dash line for the consistent pion-nucleon-delta coupling with identical form factors.
Decreasing the cut-off in (\ref{vacuumff}) to $\Lambda=0.93$GeV and increasing the off-shell one in
(\ref{medff}) to $\Lambda_\pi=1.45$GeV produces the dash-dot line for consistent coupling. 
Incorporating the form factor (\ref{ffadd}) brings the dash line to coincide with the solid one. 
\label{crossfig}}
\end{figure}

As an observable of special interest regarding the isobar in-medium properties the nuclear
photoabsorption deserves special attention. Our main purpose here is the comparison of the
isobar s-channel contribution (to the photon-nucleon scattering amplitude) when the dressing
of the isobar is performed by consistent and inconsistent couplings, thus we refrain from 
discussion of background contributions \cite{RLK-2009}.
The total nuclear photoabsorption cross section per nucleon can be written as \cite{KD-2004}:
\begin{equation}
\sigma_T(k,q)=\frac{1}{2q\cdot k}{\text{Im}}\,A_{\gamma N}(q,k),
\label{crossmed}
\end{equation}
where $q$ and $k$ are the photon and nucleon four-momenta and $A_{\gamma N}(q,k)$ is the photon-nucleon
forward scattering amplitude. Taking for the $\gamma N\Delta$ interaction the dominant magnetic dipole
form \cite{PP-2003} which is of consistent type, the forward scattering amplitude based on the tree level 
amplitude involving the s-channel isobar is given by \cite{KD-2004}:
\begin{eqnarray}
A_{\gamma\text{N}}(q,k)&=&\frac{1}{3}\,g_m^2\,h_m^2\,\left(
\sum_{i,j=1}^2 a^{(Q)}_{[ij]}(k,q) G^{(Q)}_{[ij]}(q+k,u) \right. \nonumber \\
&+& \left. 
\sum_{i,j=1}^6 a^{(P)}_{[ij]}(k,q) G^{(P)}_{[ij]}(q+k,u) \right),
\label{gnamp}
\end{eqnarray}
where
$h_m\equiv 3e/2M_N(M_N+M_\Delta)$
and $g_m$ is the magnetic dipole coupling. The expressions for nonzero $a^{(Q)}_{[ij]}(k,q)$ and 
$a^{(P)}_{[ij]}(k,q)$ terms are 
given in the Appendix by expressions (\ref{as}) for the nucleon with mean-field energy and mass shift. 
In Fig.~\ref{crossfig} the 
nuclear photoabsorption cross section is given as calculated with isobar dressed using inconsistent and 
consistent pion-nucleon-isobar couplings. The full line gives the result with inconsistent coupling where an
in-medium mass shift of $-80$MeV for the isobar was introduced as a compensation for the nucleon binding.
Performing the computation with the consistent coupling produces the dash line. The difference between the two
results can be attributed to the higher derivative order of the consistent coupling leading to sharper and
somewhat softer spectral functions as one can see in Figs.~\ref{medprop-1} and \ref{medprop-2}. Compensating the 
higher momentum power by the additional form factor (\ref{ffadd}) brings the consistent-coupling result 
to coincide with that of the inconsistent one. A similar effect can be achieved by decreasing the cut-off 
in (\ref{vacuumff}) to $\Lambda=0.93$GeV (still giving a reasonable pion-nucleon phaseshift), increasing the 
off-shell one in (\ref{medff}) to $\Lambda_\pi=1.45$GeV and increasing the isobar in-medium mass by 10MeV, 
for which case the consistent-coupling result is shown in Fig.~\ref{crossfig}
by the dash-dot line.
We observe that that the most significant contributions to the cross section come from 
$G^{(Q)}_{[11]}(p,u)$ and $G^{(P)}_{[55]}(p,u)$, i.e.\ the dominant helicity-3/2 and 
helicity-1/2 terms stemming from (positive energy) vacuum spin-3/2 sector.

\section{Conclusions \label{s3}}
We have studied the effect of using consistent and inconsistent $\pi\text{N}\Delta$ couplings
for isobar dressing in vacuum and nuclear medium. Using a convenient basis we performed complete
relativistic calculation of the isobar self energy based on the pion-nucleon loop 
for both consistent and inconsistent couplings and then 
computed the complete relativistic propagator. The vacuum computation confirmed that the 
inconsistent coupling induced new structure in the spin-1/2 sector of the vacuum propagator which was absent
when using the consistent interaction. The vacuum self energy components of the spin-3/2
sector show identical structure for both couplings, as one can see by comparing  
the $Q'$-space terms of (\ref{incvacself})
and (\ref{convacself}), apart from the $p^2$ factor for the consistent coupling coming 
from its higher
derivative nature. The spin-1/2 terms induced by the inconsistent coupling are about two orders of
magnitude smaller than the (positive energy) spin-3/2 term if the off-shell parameter $a$ in expression
(\ref{pindinc}) is of order one, making their contribution to observables receiving contribution also 
from the spin-3/2 sector negligible. However, these terms may play essential role in affecting the 
spin-1/2 background, thus obscuring the origin of different contributions to the latter. Apart from clearly
separating the origin of different spin contributions the advantage of the consistent coupling is the
much simpler structure of the isobar self energy contributions in a suitable basis as the one used in 
the present approach.

One apparent disadvantage of consistent interactions is that they are in general of higher derivative order than
the inconsistent ones \cite{Vrancx-2011a}. This means that suppression of high-momentum contributions
needs some adjustments compared to the case with inconsistent coupling. However, as shown by direct calculation, 
the difference is straightforward to establish and simple to make the needed correction
in the applied form factor. Only if a specific subtraction scheme relying on definite high-momentum 
behavior is used instead of form factors, as in 
Ref.~\cite{RLK-2009}, would the introduction of higher-momentum terms necessitate a possibly significant 
revision of the subtraction procedure. 

In the nuclear medium, even if it is spherically symmetric, the spin quantum number of a moving particle 
loses its meaning and only the helicity is a good quantum number \cite{KL-2003,KD-2004}. The isobar self 
energy acquires considerably more terms depending on both the energy and momentum, but the simpler structure
for the consistent coupling is still present as one can see by comparing expressions (\ref{conselfmed}) and
(\ref{incselfmed}).  The expressions responsible for the dominant terms in the isobar propagator,
$\sigma^{(Q)}_{[11]}(p,u)$ and $\sigma^{(P)}_{[55]}(p,u)$, are the same for both couplings apart from
the ubiquitous factor of $p^2$ for the higher-derivative consistent coupling. As a consequence the dominant 
helicity-3/2 and helicity-1/2 spectral functions differ slightly if the same form factor is applied 
as shown in Figs.~\ref{medprop-1} and \ref{medprop-2}. This difference can be easily eliminated by inserting
a supplemental form-factor compensating the additional momentum factor in the consistent coupling or by adjusting 
the cut-off parameter in the original form factor. 
Calculation of the nuclear photoabsorption cross section based on the magnetic dipole coupling which is
of consistent type reflects the similarity in the $G^{(Q)}_{[11]}(p,u)$ and $G^{(P)}_{[55]}(p,u)$ spectral
functions since the other terms in decomposition (\ref{medprop}) are much smaller and do not give significant  
contributions to the cross section in the present calculation, but this may change in case of larger
densities or other observables. 

Based on the present explicit calculation we conclude that using the consistent pion-nucleon-isobar coupling
has a number of advantages compared to the widely used inconsistent form. In case of free space (vacuum) 
computations it is primarily the elimination of the relatively small (compared to the spin-3/2 term) 
spurious spin-1/2 background of
the induced spin-1/2 sector of the isobar propagator but also the much simpler form (in a suitable basis)
of the isobar self energy and dressed propagator. 
Simplification of the full isobar self energy is also present in the nuclear medium although in this case
the spin is not a good quantum number anymore and even the consistent coupling allows for mixing of
different spin values.  
However, a possible ambiguity connected to
different off-shell parameter values in (\ref{pindinc}) is avoided by using the consistent interaction. 
The presence of higher powers of momentum in vacuum and in-medium isobar
self energy when using the consistent coupling requires some adjustment of the used form factors and possibly the isobar 
in-medium mass when fitting physical observables, but the required correction is straightforward to perform.

\appendix*
\section{}
The pion-nucleon-loop isobar vacuum self energy was calculated in Ref.\ \cite{Korpa-1997fk} 
for the inconsistent coupling (\ref{pindinc}), but since there a different basis was used we give here
the nonzero coefficients of the decomposition (\ref{vacself}):
\begin{eqnarray}
\sigma^{(Q')}_{V[11]} && =\beta_1+\sqrt{p^2}\beta_2,\nonumber\\ 
\sigma^{(Q')}_{V[22]} && =\beta_1-\sqrt{p^2}\beta_2,\nonumber\\ 
\sigma^{(P)}_{V[11]} && =\beta_1-\sqrt{p^2}\beta_2+3(\beta_5-\beta_6),\nonumber\\ 
\sigma^{(P)}_{V[22]} && =\beta_1+\sqrt{p^2}\beta_2+3(\beta_5+\beta_6),\nonumber\\ 
\sigma^{(P)}_{V[33]} && =\beta_1+\sqrt{p^2}\beta_2+p^2(\beta_3+\sqrt{p^2}\beta_4)
+\beta_5 \nonumber \\ 
&&+\beta_6+p^2(\beta_7+\beta_8)+\sqrt{p^2}(\beta_9+\beta_{10}),\nonumber \\ 
\sigma^{(P)}_{V[44]} && =\beta_1-\sqrt{p^2}\beta_2+p^2(\beta_3-\sqrt{p^2}\beta_4)
+\beta_5\nonumber\\ 
&&-\beta_6+p^2(\beta_7+\beta_8)-\sqrt{p^2}(\beta_9+\beta_{10}),\nonumber\\ 
\sigma^{(P)}_{V[13]} && =\sigma^{(P)}_{V[31]}=\sqrt{3}(\beta_5+\beta_6+p^2\beta_8+\sqrt{p^2}
\beta_{10}),\nonumber\\ 
\sigma^{(P)}_{V[24]} && =\sigma^{(P)}_{V[42]}=\sqrt{3}(-\beta_5+\beta_6-p^2\beta_8+\sqrt{p^2}
\beta_{10}),  \label{incvacself}
\end{eqnarray}
with $\beta_i\equiv\beta_i(p^2)$ given in Ref.\ \cite{Korpa-1997fk} with coupling $g$ corresponding
to $g_{\pi\text{N}\Delta}$.

For the consistent coupling (\ref{pindcon}) the number of nonzero terms is much smaller 
and we see the effect of higher-derivative coupling in the additional factor of $p^2$
(the coupling
$g$ should now be replaced with $g^{\text{(con)}}_{\pi\text{N}\Delta}/M_\Delta$):
\begin{eqnarray}
\sigma^{(Q')}_{V[11]} &=& p^2(\beta_1+\sqrt{p^2}\beta_2),\nonumber\\ 
\sigma^{(Q')}_{V[22]} &=& p^2(\beta_1-\sqrt{p^2}\beta_2),\nonumber\\ 
\sigma^{(P)}_{V[11]} &=& 4p^2(\beta_1-\sqrt{p^2}\beta_2),\nonumber\\ 
\sigma^{(P)}_{V[22]} &=& 4p^2(\beta_1+\sqrt{p^2}\beta_2). \label{convacself}
\end{eqnarray}
 
The in-medium isobar propagator depends also on the 4-vector $u$ characterizing the medium 4-velocity.
We present the nonzero expressions for the isobar self-energy expansion coefficients $\sigma^{(Q,P)}_{[ij]}$ 
in (\ref{medself}) showing only term with $i\leq j$ since they are symmetric under interchange of $i$ and $j$.
We start with the case of consistent coupling (\ref{pindcon}):
\begin{eqnarray}
&&\sigma^{(Q)}_{[11]}=(M^*_N +\Sigma_N^v \hat p_0)L_3+L_7,\;\;
\sigma^{(Q)}_{[12]}=i (L_8-\Sigma_N^v \hat p_v L_3),\nonumber \\
&&\sigma^{(Q)}_{[22]}=(M^*_N -\Sigma_N^v \hat p_0)L_3-L_7,\nonumber \\
&&\sigma^{(P)}_{[11]}=\frac{4}{3}\left[ (M^*_N -\Sigma_N^v \hat p_0)(2L_3-L_5)-2L_7+L_{12}\right],\nonumber\\
&&\sigma^{(P)}_{[12]}=-\frac{4i}{3}\left( 3L_8+L_{11}-\Sigma_N^v \hat p_v(2L_3+L_5)\right),\nonumber\\
&&\sigma^{(P)}_{[14]}=i\sqrt{3}L_8\nonumber\\
&&\sigma^{(P)}_{[15]}=\frac{2\sqrt{2}i}{3}\left(2L_8-L_{11}-\Sigma_N^v \hat p_v(L_3-L_5)\right),\nonumber\\
&&\sigma^{(P)}_{[16]}=-\frac{2\sqrt{2}}{3}\left[(M^*_N -\Sigma_N^v \hat p_0)(L_3+L_5)-L_7-L_{12}\right],\nonumber\\
&&\sigma^{(P)}_{[22]}=\frac{4}{3}\left[ (M^*_N +\Sigma_N^v \hat p_0)(2L_3-L_5)+2L_7-L_{12}\right],\nonumber \\
&&\sigma^{(P)}_{[23]}=-i\sqrt{3}L_8\nonumber\\
&&\sigma^{(P)}_{[25]}=-\frac{2\sqrt{2}}{3}\left[(M^*_N +\Sigma_N^v \hat p_0)(L_3+L_5)+L_7+L_{12}\right],
\nonumber\\
&&\sigma^{(P)}_{[26]}=\frac{2\sqrt{2}i}{3}\left(2L_8-L_{11}-\Sigma_N^v \hat p_v(L_3-L_5)\right),
\nonumber\\
&&\sigma^{(P)}_{[55]}=\frac{1}{3}\left[(M^*_N +\Sigma_N^v \hat p_0)(L_3-2L_5)+L_7-2L_{12}\right],\nonumber\\
&&\sigma^{(P)}_{[56]}=\frac{i}{3}\left(\Sigma_N^v \hat p_v(L_3+2L_5)-5L_8-2L_{11}\right),\nonumber\\
&&\sigma^{(P)}_{[66]}=\frac{1}{3}\left[(M^*_N -\Sigma_N^v \hat p_0)(L_3-2L_5)-L_7+2L_{12}\right].
\label{conselfmed}
\end{eqnarray}
Here $\Sigma_N^s$ and $\Sigma_N^v$ are the scalar and vector parts of the mean-field nucleon self energy, with 
$M^*_N\equiv M_N+\Sigma_N^s$.

For the case of inconsistent coupling (\ref{pindinc}) we give the coefficients $\sigma^{(Q,P)}_{[ij]}$ for
the case with $a=0$, but generalizing the expressions given in Ref.~\cite{KD-2004} to the case with 
nucleon mean-field self-energy shifts taken into account:
\begin{widetext}
\begin{eqnarray}
&&\sigma^{(Q)}_{[11]}=(M^*_N +\Sigma_N^v \hat p_0)L_3+L_7,\;\;
\sigma^{(Q)}_{[12]}=-i (L_8-\Sigma_N^v \hat p_v L_3),\nonumber \\
&&\sigma^{(Q)}_{[22]}=(M^*_N -\Sigma_N^v \hat p_0)L_3-L_7,\nonumber \\
&&\sigma^{(P)}_{[11]}=\frac{1}{3}\left[ (M^*_N -\Sigma_N^v \hat p_0)(2L_3-L_5)-2L_7+L_{12}\right],\nonumber\\
&&\sigma^{(P)}_{[12]}=\frac{i}{3}\left[ -2L_8+L_{11}-\Sigma_N^v \hat p_v (2L_3+L_5)\right],\nonumber\\
&&\sigma^{(P)}_{[13]}=\frac{1}{\sqrt{3}}\left[-\sqrt{p^2}(2L_3-L_5)+2L_7-L_{12}+\Sigma_N^v \hat p_v
(L_6-\sqrt{p^2}L_2)\right],\nonumber\\
&&\sigma^{(P)}_{[14]}=\frac{i}{\sqrt{3}}\left[M_N(\sqrt{p^2}L_2-L_6)-\sqrt{p^2}L_6+%
L_{10}+\Sigma_N^v \hat p_v(L_6-\sqrt{p^2}L_2)\right],\nonumber\\
&&\sigma^{(P)}_{[15]}=\frac{i\sqrt{2}}{3}\left[2L_8-L_{11}+\Sigma_N^v \hat p_v(L_5-L_3)\right],\nonumber\\
&&\sigma^{(P)}_{[16]}=\frac{\sqrt{2}}{3}\left[(M^*_N -\Sigma_N^v \hat p_0)(L_3+L_5)-L_7-L_{12}\right],\nonumber\\
&&\sigma^{(P)}_{[22]}=\frac{1}{3}\left[(M^*_N +\Sigma_N^v \hat p_0)(2L_3-L_5)+2L_7-L_{12}\right],\nonumber\\
&&\sigma^{(P)}_{[23]}=\frac{i}{\sqrt{3}}\left[(M^*_N +\Sigma_N^v \hat p_0)(\sqrt{p^2}L_2-L_6)-\sqrt{p^2}L_6-%
L_{10}\right],\nonumber\\
&&\sigma^{(P)}_{[24]}=\frac{1}{\sqrt{3}}\left[-\sqrt{p^2}(2L_3-L_5)+2L_7-L_{12}+\Sigma_N^v \hat p_v(L_6
-\sqrt{p^2}L_2)\right],
\nonumber\\
&&\sigma^{(P)}_{[25]}=\frac{\sqrt{2}}{3}\left[(M^*_N +\Sigma_N^v \hat p_0)(L_3+L_5)+L_7+L_{12}\right],\nonumber\\
&&\sigma^{(P)}_{[26]}=\frac{i\sqrt{2}}{3}\left[2L_8-L_{11}+\sigma_N^v\hat p_v(L_5-L_3)\right],
\nonumber\\
&&\sigma^{(P)}_{[33]}=(M^*_N +\Sigma_N^v \hat p_0) p^2 L_0+(p^2-2(M^*_N +\Sigma_N^v \hat p_0)\sqrt{p^2})L_1
+(M^*_N +\Sigma_N^v \hat p_0-2\sqrt{p^2})L_4
+L_9,\nonumber\\
&&\sigma^{(P)}_{[34]}=i\left[-p^2 L_2+2\sqrt{p^2}L_6-L_{10}+\sigma_N^v\hat p_v (p^2 L_0-2\sqrt{p^2}L_1+
L_4)\right],\nonumber\\
&&\sigma^{(P)}_{[35]}=i\sqrt{\frac{2}{3}}\left[ -(M^*_N +\Sigma_N^v \hat p_0)\sqrt{p^2} L_2+
(M^*_N +\Sigma_N^v \hat p_0-\sqrt{p^2})L_6+L_{10}\right],\nonumber\\
&&\sigma^{(P)}_{[36]}=\sqrt{\frac{2}{3}}\left[ -\sqrt{p^2} (L_3+L_5)
+L_7+L_{12}+\Sigma_N^v \hat p_v(\sqrt{p^2}L_2-L_6)\right],\nonumber\\
&&\sigma^{(P)}_{[44]}=(M^*_N -\Sigma_N^v \hat p_0) p^2 L_0-(p^2+2(M^*_N -\Sigma_N^v \hat p_0)\sqrt{p^2})L_1
+(M^*_N -\Sigma_N^v \hat p_0-2\sqrt{p^2})L_4
+L_9,\nonumber\\
&&\sigma^{(P)}_{[45]}=\sqrt{\frac{2}{3}} \left[ -\sqrt{p^2}(L_3+L_5)+L_7+L_{12}+\Sigma_N^v \hat p_v 
(\sqrt{p^2}L_2-L_6)\right],
\nonumber\\
&&\sigma^{(P)}_{[46]}=i\sqrt{\frac{2}{3}} \left[-(M^*_N -\Sigma_N^v \hat p_0) \sqrt{p^2} L_2+
(M^*_N -\Sigma_N^v \hat p_0+\sqrt{p^2})L_6-L_{10}\right],\nonumber\\
&&\sigma^{(P)}_{[55]}=\frac{1}{3}\left[(M^*_N +\Sigma_N^v \hat p_0)(L_3-2L_5)+L_7-2L_{12}\right],\nonumber\\
&&\sigma^{(P)}_{[56]}=\frac{i}{3}\left[ 5L_8+2L_{11}-\Sigma_N^v \hat p_v (L_3+2L_5)\right],\nonumber\\
&&\sigma^{(P)}_{[66]}=\frac{1}{3}\left[(M^*_N -\Sigma_N^v \hat p_0)(L_3-2L_5)-L_7+2L_{12}\right],
\label{incselfmed}
\end{eqnarray}
with $\hat p_0\equiv p_0/\sqrt{p^2}$, $\hat p_v \equiv |\vec p|/\sqrt{p^2}$.
The pion-nucleon loop integrals $L_i(p,u)$ are regulated by the form factor in the pion-nucleon-isobar
vertex. That assures the imaginary part of the loop approaches zero at large energy sufficiently fast
for the dispersion integral of the real part to be convergent. The imaginary part of the loop 
integrals is given by \cite{KD-2004}:
\begin{equation}
\text{Im}\, L_i(p,u)=\frac{g^2}{8\pi^2}
\int_{k_F}^\infty \frac{\vec k\,^2 dk}{E_k}\int_{-1}^1 d\mu \,
F(p,k)^2\; \text{Im}\,D_\pi(p_0-E_k,|\vec p - \vec k|) K_i(p,k),
\end{equation}
where $F(p,k)$ is the $\pi\textrm{N}\Delta$ form factor, 
$\mu\equiv \cos\theta(\vec p,\vec k)$, $D_\pi$ the in-medium pion propagator 
and the functions $K_i(p,k)$ are defined in Ref.~\cite{KD-2004}. In connection with
the latter we remark that in $K_3(p,k)$ the $M^2_N$ should be replaced by $k_0^2-\vec k\,^2$ where
$k_0=\sqrt{M^{*2}_N+\vec k\,^2}+\Sigma_N^v$.
The coupling
$g$ denotes $g_{\pi\text{N}\Delta}$ for the case of inconsistent pion-nucleon-isobar
interaction (\ref{pindinc}), while for the consistent interaction (\ref{pindcon}) it coresponds to
$\sqrt{p^2}\,g^{\text{(con)}}_{\pi\text{N}\Delta}/M_\Delta$.

The functions $a^{(Q)}_{[ij]}(k,q)$ and $a^{(P)}_{[ij]}(k,q)$ are calculated for the nucleon with
mean-field scalar and vector self energies thus again generalizing corresponding results of Ref.~\cite{KD-2004}:
\begin{eqnarray}
a^{(Q)}_{[11]} &=& \frac{p^2}{2}(M^*_++\Sigma_N^v \hat p_0)\left[ (X\cdot k)^2-k^2+(k\cdot \hat p)^2 \right],
\nonumber\\
a^{(Q)}_{[12]} &=& \frac{-i p^2}{2}(X\cdot k^* +\Sigma_N^v \hat p_v) \left[ k^2 -(k\cdot \hat p)^2-(X\cdot k)^2\right],
  \nonumber\\
a^{(Q)}_{[22]} &=& \frac{p^2}{2}(M^*_--\Sigma_N^v \hat p_0)\left[ (X\cdot k)^2-k^2+(k\cdot \hat p)^2 \right],
\nonumber\\
a^{(P)}_{[11]} &=& \frac{-2p^2}{3}(M^*_--\Sigma_N^v \hat p_0)\left(k^2-(k\cdot \hat p)^2 \right),
\nonumber\\
a^{(P)}_{[12]} &=& \frac{4ip^2}{3}X\cdot k \,(k\cdot k^* -k\cdot\hat p\,\,k^*\cdot \hat p -\Sigma_N^v \hat p_v\,
X\cdot k),\nonumber\\
a^{(P)}_{[15]} &=& \frac{i\sqrt{2}p^2}{3}X\cdot k\, \left[ 3k^2-k\cdot k^*\, -3(k\cdot \hat p)^2+k\cdot\hat p\,
k^*\cdot\hat p+\Sigma_N^v \hat k_v X\cdot k\right],\nonumber\\
a^{(P)}_{[16]} &=& \frac{\sqrt{2}p^2}{3}(M^*_--\Sigma_N^v \hat p_0) \left[ k^2-(k\cdot\hat p)^2+3\left( X\cdot k\right)^2
\right],\nonumber \\
a^{(P)}_{[22]} &=& \frac{-2p^2}{3}(M^*_++\Sigma_N^v \hat p_0)\left(k^2-(k\cdot \hat p)^2 \right),
\nonumber\\
a^{(P)}_{[25]} &=& \frac{\sqrt{2}p^2}{3}(M^*_++\Sigma_N^v \hat p_0) \left[ k^2-(k\cdot\hat p)^2+3\left( X\cdot k\right)^2
\right],\nonumber\\
a^{(P)}_{[26]} &=& a^{(P)}_{15},\nonumber\\
a^{(P)}_{[55]} &=& \frac{-p^2}{6}(M^*_++\Sigma_N^v \hat p_0) \left[5k^2-5(k\cdot\hat p)^2+3\left( X\cdot k\right)^2
\right],\nonumber\\
a^{(P)}_{[56]} &=& \frac{ip^2}{3}\left[ -4X\cdot k\,(k\cdot k^* -k\cdot\hat p\, k^*\cdot\hat p)+(\Sigma_N^v \hat p_v+
X\cdot k^*)((k\cdot\hat p)^2-k^2)\right. \nonumber \\
&&\left. -(X\cdot k)^2(5\Sigma_N^v \hat p_v+9X\cdot k^*)
\right],\nonumber\\
a^{(P)}_{[66]} &=& \frac{-p^2}{6}(M^*_--\Sigma_N^v \hat p_0) \left[5k^2-5(k\cdot\hat p)^2+3\left( X\cdot k\right)^2
\right],
\label{as}
\end{eqnarray}
\end{widetext}
where $M^*_\pm\equiv M^*_N \pm k^*\cdot p/\sqrt{p^2}$, $\hat p\equiv p/\sqrt{p^2}$ and 
$k^*\equiv (\sqrt{M_N^{*2}+\vec k\,^2},\vec k\,)$.

\begin{acknowledgments}
This research was supported in part by the Hungarian Research Foundation (OTKA)
grants 71989 and T48833.
\end{acknowledgments}

\bibliography{paper}

\providecommand{\noopsort}[1]{}\providecommand{\singleletter}[1]{#1}%
\begin{thebibliography}{19}%
\makeatletter
\providecommand \@ifxundefined [1]{%
 \@ifx{#1\undefined}
}%
\providecommand \@ifnum [1]{%
 \ifnum #1\expandafter \@firstoftwo
 \else \expandafter \@secondoftwo
 \fi
}%
\providecommand \@ifx [1]{%
 \ifx #1\expandafter \@firstoftwo
 \else \expandafter \@secondoftwo
 \fi
}%
\providecommand \natexlab [1]{#1}%
\providecommand \enquote  [1]{``#1''}%
\providecommand \bibnamefont  [1]{#1}%
\providecommand \bibfnamefont [1]{#1}%
\providecommand \citenamefont [1]{#1}%
\providecommand \href@noop [0]{\@secondoftwo}%
\providecommand \href [0]{\begingroup \@sanitize@url \@href}%
\providecommand \@href[1]{\@@startlink{#1}\@@href}%
\providecommand \@@href[1]{\endgroup#1\@@endlink}%
\providecommand \@sanitize@url [0]{\catcode `\\12\catcode `\$12\catcode
  `\&12\catcode `\#12\catcode `\^12\catcode `\_12\catcode `\%12\relax}%
\providecommand \@@startlink[1]{}%
\providecommand \@@endlink[0]{}%
\providecommand \url  [0]{\begingroup\@sanitize@url \@url }%
\providecommand \@url [1]{\endgroup\@href {#1}{\urlprefix }}%
\providecommand \urlprefix  [0]{URL }%
\providecommand \Eprint [0]{\href }%
\providecommand \doibase [0]{http://dx.doi.org/}%
\providecommand \selectlanguage [0]{\@gobble}%
\providecommand \bibinfo  [0]{\@secondoftwo}%
\providecommand \bibfield  [0]{\@secondoftwo}%
\providecommand \translation [1]{[#1]}%
\providecommand \BibitemOpen [0]{}%
\providecommand \bibitemStop [0]{}%
\providecommand \bibitemNoStop [0]{.\EOS\space}%
\providecommand \EOS [0]{\spacefactor3000\relax}%
\providecommand \BibitemShut  [1]{\csname bibitem#1\endcsname}%
\let\auto@bib@innerbib\@empty
\bibitem [{\citenamefont {Rarita}\ and\ \citenamefont
  {Schwinger}(1941)}]{RS-1941}%
  \BibitemOpen
  \bibfield  {author} {\bibinfo {author} {\bibfnamefont {W.}~\bibnamefont
  {Rarita}}\ and\ \bibinfo {author} {\bibfnamefont {J.}~\bibnamefont
  {Schwinger}},\ }\href@noop {} {\bibfield  {journal} {\bibinfo  {journal}
  {Phys.\ Rev.}\ }\textbf {\bibinfo {volume} {60}},\ \bibinfo {pages} {61}
  (\bibinfo {year} {1941})}\BibitemShut {NoStop}%
\bibitem [{\citenamefont {Pascalutsa}(1998)}]{Pascalutsa-1998}%
  \BibitemOpen
  \bibfield  {author} {\bibinfo {author} {\bibfnamefont {V.}~\bibnamefont
  {Pascalutsa}},\ }\href@noop {} {\bibfield  {journal} {\bibinfo  {journal}
  {Phys.\ Rev.\ D}\ }\textbf {\bibinfo {volume} {58}},\ \bibinfo {pages}
  {096002} (\bibinfo {year} {1998})}\BibitemShut {NoStop}%
\bibitem [{\citenamefont {Pascalutsa}\ and\ \citenamefont
  {Timmermans}(1999)}]{PT-1999}%
  \BibitemOpen
  \bibfield  {author} {\bibinfo {author} {\bibfnamefont {V.}~\bibnamefont
  {Pascalutsa}}\ and\ \bibinfo {author} {\bibfnamefont {R.}~\bibnamefont
  {Timmermans}},\ }\href@noop {} {\bibfield  {journal} {\bibinfo  {journal}
  {Phys.\ Rev.\ C}\ }\textbf {\bibinfo {volume} {60}},\ \bibinfo {pages}
  {042201} (\bibinfo {year} {1999})}\BibitemShut {NoStop}%
\bibitem [{\citenamefont {Pascalutsa}(2001)}]{Pascalutsa-2001}%
  \BibitemOpen
  \bibfield  {author} {\bibinfo {author} {\bibfnamefont {V.}~\bibnamefont
  {Pascalutsa}},\ }\href@noop {} {\bibfield  {journal} {\bibinfo  {journal}
  {Phys.\ Lett.\ B}\ }\textbf {\bibinfo {volume} {503}},\ \bibinfo {pages} {85}
  (\bibinfo {year} {2001})}\BibitemShut {NoStop}%
\bibitem [{\citenamefont {Pascalutsa}(2008)}]{Pascalutsa-2008}%
  \BibitemOpen
  \bibfield  {author} {\bibinfo {author} {\bibfnamefont {V.}~\bibnamefont
  {Pascalutsa}},\ }\href@noop {} {\bibfield  {journal} {\bibinfo  {journal}
  {Prog.\ Part.\ Nucl.\ Phys.}\ }\textbf {\bibinfo {volume} {61}},\ \bibinfo
  {pages} {27} (\bibinfo {year} {2008})}\BibitemShut {NoStop}%
\bibitem [{\citenamefont {Vrancx}\ \emph
  {et~al.}(2011{\natexlab{a}})\citenamefont {Vrancx}, \citenamefont {Cruz},
  \citenamefont {Ryckebusch},\ and\ \citenamefont
  {Vancraeyveld}}]{Vrancx-2011a}%
  \BibitemOpen
  \bibfield  {author} {\bibinfo {author} {\bibfnamefont {T.}~\bibnamefont
  {Vrancx}}, \bibinfo {author} {\bibfnamefont {L.~D.}\ \bibnamefont {Cruz}},
  \bibinfo {author} {\bibfnamefont {J.}~\bibnamefont {Ryckebusch}}, \ and\
  \bibinfo {author} {\bibfnamefont {P.}~\bibnamefont {Vancraeyveld}},\
  }\href@noop {} {} (\bibinfo {year} {2011}{\natexlab{a}}),\ \Eprint
  {http://arxiv.org/abs/arXiv/1105.0780v1 [nucl-th]} {arXiv/1105.0780v1
  [nucl-th]} \BibitemShut {NoStop}%
\bibitem [{\citenamefont {Vrancx}\ \emph
  {et~al.}(2011{\natexlab{b}})\citenamefont {Vrancx}, \citenamefont {Cruz},
  \citenamefont {Ryckebusch},\ and\ \citenamefont
  {Vancraeyveld}}]{Vrancx-2011b}%
  \BibitemOpen
  \bibfield  {author} {\bibinfo {author} {\bibfnamefont {T.}~\bibnamefont
  {Vrancx}}, \bibinfo {author} {\bibfnamefont {L.~D.}\ \bibnamefont {Cruz}},
  \bibinfo {author} {\bibfnamefont {J.}~\bibnamefont {Ryckebusch}}, \ and\
  \bibinfo {author} {\bibfnamefont {P.}~\bibnamefont {Vancraeyveld}},\
  }\href@noop {} {} (\bibinfo {year} {2011}{\natexlab{b}}),\ \Eprint
  {http://arxiv.org/abs/arXiv/1108.2688v1 [nucl-th]} {arXiv/1108.2688v1
  [nucl-th]} \BibitemShut {NoStop}%
\bibitem [{\citenamefont {Nath}\ \emph {et~al.}(1971)\citenamefont {Nath},
  \citenamefont {Etemadi},\ and\ \citenamefont {Kimel}}]{NEK-1971}%
  \BibitemOpen
  \bibfield  {author} {\bibinfo {author} {\bibfnamefont {L.~M.}\ \bibnamefont
  {Nath}}, \bibinfo {author} {\bibfnamefont {B.}~\bibnamefont {Etemadi}}, \
  and\ \bibinfo {author} {\bibfnamefont {J.~D.}\ \bibnamefont {Kimel}},\
  }\href@noop {} {\bibfield  {journal} {\bibinfo  {journal} {Phys.\ Rev.\ D}\
  }\textbf {\bibinfo {volume} {3}},\ \bibinfo {pages} {2153} (\bibinfo {year}
  {1971})}\BibitemShut {NoStop}%
\bibitem [{\citenamefont {Lutz}\ and\ \citenamefont {Korpa}(2002)}]{LK-2002}%
  \BibitemOpen
  \bibfield  {author} {\bibinfo {author} {\bibfnamefont {M.~F.~M.}\
  \bibnamefont {Lutz}}\ and\ \bibinfo {author} {\bibfnamefont {C.~L.}\
  \bibnamefont {Korpa}},\ }\href@noop {} {\bibfield  {journal} {\bibinfo
  {journal} {Nucl.\ Phys.\ A}\ }\textbf {\bibinfo {volume} {700}},\ \bibinfo
  {pages} {309} (\bibinfo {year} {2002})}\BibitemShut {NoStop}%
\bibitem [{\citenamefont {Korpa}\ and\ \citenamefont
  {Dieperink}(2004)}]{KD-2004}%
  \BibitemOpen
  \bibfield  {author} {\bibinfo {author} {\bibfnamefont {C.~L.}\ \bibnamefont
  {Korpa}}\ and\ \bibinfo {author} {\bibfnamefont {A.}~\bibnamefont
  {Dieperink}},\ }\href@noop {} {\bibfield  {journal} {\bibinfo  {journal}
  {Phys.\ Rev.\ C}\ }\textbf {\bibinfo {volume} {70}},\ \bibinfo {pages}
  {015207} (\bibinfo {year} {2004})}\BibitemShut {NoStop}%
\bibitem [{\citenamefont {Korpa}(1997)}]{Korpa-1997fk}%
  \BibitemOpen
  \bibfield  {author} {\bibinfo {author} {\bibfnamefont {C.~L.}\ \bibnamefont
  {Korpa}},\ }\href@noop {} {\bibfield  {journal} {\bibinfo  {journal} {Heavy
  Ion Phys.}\ }\textbf {\bibinfo {volume} {5}},\ \bibinfo {pages} {77}
  (\bibinfo {year} {1997})},\ \bibinfo {note} {[Erratum-ibid.\
  5:319-320,1997]}\BibitemShut {NoStop}%
\bibitem [{\citenamefont {Kaloshin}\ and\ \citenamefont
  {Lomov}(2006)}]{KL-2006}%
  \BibitemOpen
  \bibfield  {author} {\bibinfo {author} {\bibfnamefont {A.~E.}\ \bibnamefont
  {Kaloshin}}\ and\ \bibinfo {author} {\bibfnamefont {V.~P.}\ \bibnamefont
  {Lomov}},\ }\href@noop {} {\bibfield  {journal} {\bibinfo  {journal} {Phys.\
  Atom.\ Nucl.}\ }\textbf {\bibinfo {volume} {69}},\ \bibinfo {pages} {541}
  (\bibinfo {year} {2006})}\BibitemShut {NoStop}%
\bibitem [{\citenamefont {Benmerrouche}\ \emph {et~al.}(1989)\citenamefont
  {Benmerrouche}, \citenamefont {Davidson},\ and\ \citenamefont
  {Mukhopadhyay}}]{BDM-1989}%
  \BibitemOpen
  \bibfield  {author} {\bibinfo {author} {\bibfnamefont {M.}~\bibnamefont
  {Benmerrouche}}, \bibinfo {author} {\bibfnamefont {R.}~\bibnamefont
  {Davidson}}, \ and\ \bibinfo {author} {\bibfnamefont {C.}~\bibnamefont
  {Mukhopadhyay}},\ }\href@noop {} {\bibfield  {journal} {\bibinfo  {journal}
  {Phys.\ Rev.\ C}\ }\textbf {\bibinfo {volume} {39}},\ \bibinfo {pages} {2339}
  (\bibinfo {year} {1989})}\BibitemShut {NoStop}%
\bibitem [{\citenamefont {Koch}\ and\ \citenamefont
  {Pietarinen}(1980)}]{KP-1980}%
  \BibitemOpen
  \bibfield  {author} {\bibinfo {author} {\bibfnamefont {R.}~\bibnamefont
  {Koch}}\ and\ \bibinfo {author} {\bibfnamefont {E.}~\bibnamefont
  {Pietarinen}},\ }\href@noop {} {\bibfield  {journal} {\bibinfo  {journal}
  {Nucl.\ Phys.\ A}\ }\textbf {\bibinfo {volume} {336}},\ \bibinfo {pages}
  {331} (\bibinfo {year} {1980})}\BibitemShut {NoStop}%
\bibitem [{\citenamefont {Korpa}\ and\ \citenamefont {Lutz}(2003)}]{KL-2003}%
  \BibitemOpen
  \bibfield  {author} {\bibinfo {author} {\bibfnamefont {C.~L.}\ \bibnamefont
  {Korpa}}\ and\ \bibinfo {author} {\bibfnamefont {M.}~\bibnamefont {Lutz}},\
  }\href@noop {} {\bibfield  {journal} {\bibinfo  {journal} {Heavy Ion Phys.}\
  }\textbf {\bibinfo {volume} {17}},\ \bibinfo {pages} {341} (\bibinfo {year}
  {2003})}\BibitemShut {NoStop}%
\bibitem [{\citenamefont {Korpa}\ \emph {et~al.}(2009)\citenamefont {Korpa},
  \citenamefont {Lutz},\ and\ \citenamefont {Riek}}]{KLR-2009}%
  \BibitemOpen
  \bibfield  {author} {\bibinfo {author} {\bibfnamefont {C.~L.}\ \bibnamefont
  {Korpa}}, \bibinfo {author} {\bibfnamefont {M.~F.~M.}\ \bibnamefont {Lutz}},
  \ and\ \bibinfo {author} {\bibfnamefont {F.}~\bibnamefont {Riek}},\
  }\href@noop {} {\bibfield  {journal} {\bibinfo  {journal} {Phys.\ Rev.\ C}\
  }\textbf {\bibinfo {volume} {80}},\ \bibinfo {pages} {024901} (\bibinfo
  {year} {2009})}\BibitemShut {NoStop}%
\bibitem [{\citenamefont {Riek}\ \emph {et~al.}(2009)\citenamefont {Riek},
  \citenamefont {Lutz},\ and\ \citenamefont {Korpa}}]{RLK-2009}%
  \BibitemOpen
  \bibfield  {author} {\bibinfo {author} {\bibfnamefont {F.}~\bibnamefont
  {Riek}}, \bibinfo {author} {\bibfnamefont {M.~F.~M.}\ \bibnamefont {Lutz}}, \
  and\ \bibinfo {author} {\bibfnamefont {C.~L.}\ \bibnamefont {Korpa}},\
  }\href@noop {} {\bibfield  {journal} {\bibinfo  {journal} {Phys.\ Rev.\ C}\
  }\textbf {\bibinfo {volume} {80}},\ \bibinfo {pages} {024902} (\bibinfo
  {year} {2009})}\BibitemShut {NoStop}%
\bibitem [{\citenamefont {Pascalutsa}\ and\ \citenamefont
  {Phillips}(2003)}]{PP-2003}%
  \BibitemOpen
  \bibfield  {author} {\bibinfo {author} {\bibfnamefont {V.}~\bibnamefont
  {Pascalutsa}}\ and\ \bibinfo {author} {\bibfnamefont {D.~R.}\ \bibnamefont
  {Phillips}},\ }\href@noop {} {\bibfield  {journal} {\bibinfo  {journal}
  {Phys.\ Rev.\ C}\ }\textbf {\bibinfo {volume} {67}},\ \bibinfo {pages}
  {055202} (\bibinfo {year} {2003})}\BibitemShut {NoStop}%
\bibitem [{\citenamefont {Ahrens}\ \emph {et~al.}(1984)\citenamefont {Ahrens},
  \citenamefont {Arends}, \citenamefont {Bourgeois}, \citenamefont {Carlos},
  \citenamefont {Fallou}, \citenamefont {Floss}, \citenamefont {Garganne},
  \citenamefont {Huthmacher}, \citenamefont {Kneissl}, \citenamefont {Mank},
  \citenamefont {Mecking}, \citenamefont {Ries}, \citenamefont {Stenz},\ and\
  \citenamefont {Veissiere}}]{Ahrens-1984}%
  \BibitemOpen
  \bibfield  {author} {\bibinfo {author} {\bibfnamefont {J.}~\bibnamefont
  {Ahrens}}, \bibinfo {author} {\bibfnamefont {J.}~\bibnamefont {Arends}},
  \bibinfo {author} {\bibfnamefont {P.}~\bibnamefont {Bourgeois}}, \bibinfo
  {author} {\bibfnamefont {P.}~\bibnamefont {Carlos}}, \bibinfo {author}
  {\bibfnamefont {J.~L.}\ \bibnamefont {Fallou}}, \bibinfo {author}
  {\bibfnamefont {N.}~\bibnamefont {Floss}}, \bibinfo {author} {\bibfnamefont
  {P.}~\bibnamefont {Garganne}}, \bibinfo {author} {\bibfnamefont
  {S.}~\bibnamefont {Huthmacher}}, \bibinfo {author} {\bibfnamefont
  {U.}~\bibnamefont {Kneissl}}, \bibinfo {author} {\bibfnamefont
  {G.}~\bibnamefont {Mank}}, \bibinfo {author} {\bibfnamefont {B.}~\bibnamefont
  {Mecking}}, \bibinfo {author} {\bibfnamefont {H.}~\bibnamefont {Ries}},
  \bibinfo {author} {\bibfnamefont {R.}~\bibnamefont {Stenz}}, \ and\ \bibinfo
  {author} {\bibfnamefont {A.}~\bibnamefont {Veissiere}},\ }\href@noop {}
  {\bibfield  {journal} {\bibinfo  {journal} {Phys.\ Lett.\ B}\ }\textbf
  {\bibinfo {volume} {146}},\ \bibinfo {pages} {303} (\bibinfo {year}
  {1984})}\BibitemShut {NoStop}%
\end{thebibliography}%

\end{document}